\documentclass{article}
\usepackage[utf8]{inputenc}

\usepackage{PRIMEarxiv}
\usepackage{subfig}
\usepackage[utf8]{inputenc} 
\usepackage[T1]{fontenc}    
\usepackage{hyperref}       
\usepackage{url}            
\usepackage{booktabs}       
\usepackage{amsfonts}       
\usepackage{nicefrac}       
\usepackage{microtype}      
\usepackage{lipsum}
\usepackage{fancyhdr}       
\usepackage{graphicx}       
\graphicspath{{media/}}     
\usepackage{amsmath}
\usepackage{xcolor}
\title{DiffusionInv: Prior-enhanced Bayesian Full Waveform Inversion using Diffusion models

}

\author{
  Yuanyuan Li \\
   China University of Petroleum (East China)\\
  \texttt{yuanyuanli@upc.edu.cn} \\
  \And
  Hao Zhang \\
   China University of Petroleum (East China)\\
  \texttt{zhanghao@s.upc.edu.com} \\
   \And
  Zhuoqi Yan \\
   China University of Petroleum (East China)\\
  \texttt{zhuoqiyan@s.upc.edu.cn} \\
     \And
  Tariq Alkhalifah \\
  King Abdullah University of Science and Technology\\
  \texttt{tariq.alkhalifah@kaust.edu.sa} 
}

\begin{document}
\maketitle

\begin{abstract}
Full waveform inversion (FWI) is an advanced seismic inversion technique for quantitatively estimating subsurface properties by minimizing the misfit between simulated and observed seismic data. However, with FWI, it is hard to converge to a geologically-realistic subsurface model considering the large model space and the limited seismic data coverage. Thus, we propose a DiffusionInv approach by integrating a pretrained diffusion model representation of the velocity into FWI within a Bayesian framework. We first train the diffusion model on realistic and expected velocity model samples, preferably prepared based on our geological knowledge of the subsurface  and any other source of information, like well logs, to learn their prior distribution. Once this pretraining is complete, we fine-tune the neural network parameters of the diffusion model by using an L2 norm of data misfit between simulated and observed seismic data as a loss function. This process enables the model to elegantly learn the posterior distribution of velocity models given seismic observations starting from the previously learned prior distribution. The approximate representation of the posterior probability distribution provides valuable insights into the inversion uncertainty for physics-constrained FWI given the seismic data and prior model information. Compared to regularized FWI methods, the diffusion model under a probabilistic framework allows for a more adaptable integration of prior information into FWI, thus mitigating the multi-solution issue and enhancing the reliability of the inversion result. Compared to data-driven FWI methods, the integration of physical constraints in DiffusionInv enhances its ability to generalize across diverse seismic scenarios. The test results in the Hess and Otway examples demonstrate that the DiffusionInv method is capable of recovering the velocity model with high resolution by incorporating the prior model information. The estimated posterior probability distribution helps us understand the inversion uncertainty.

\end{abstract}

\section{Introduction}

Full waveform inversion (FWI) has emerged as an advanced inversion technique for predicting subsurface properties from observed seismic data \cite{tarantola1984inversion,virieux2009overview,vigh2009developing,tromp2020seismic}. Its applications span multiple scales, including global, regional, exploration and reservoir scales, making it a versatile tool for understanding the Earth's subsurface \cite{fichtner2009full,alkhalifah2014recipe,modrak2016seismic,li2022target}. Inherently, FWI is a highly-nonlinear PDE-constrained optimization problem that inverts for the subsurface model by minimizing the misfit between simulated and observed seismic data. It includes three key ingredients, those are seismic observations in the data space, subsurface parameters in the model space, and a physical relationship between seismic data and the subsurface parameter model \cite{fichtner2010full}. Seismic data is often acquired on the Earth’s surface with limited frequency band and acquisition offset. It is challenging for FWI to converge to an optimal subsurface model, well representing realistic geological properties, considering the large model space and the limited seismic data coverage, resulting in a large Null space in which multiple parameter models can satisfy the observed data within their uncertainties \cite{rawlinson2016use}. Furthermore, FWI is a highly-nonlinear inverse problem with countless local minima complicating the convergence process \cite{alkhalifah2016full}.  

A conventional method for obtaining an estimate of the subsurface is to treat the inverse problem as a least squares optimization problem within the framework of deterministic inversion theory. Such a method typically employs a gradient-based local optimization algorithm to solve the optimization problem at an affordable computational cost \cite{martin2012stochastic}. However, without low-frequency information or a kinematically accurate starting model, this approach remains highly vulnerable to cycle-skipping and falling into a local minimum solution, limiting practical applicability in geologically complex settings.  Regularization techniques have been deployed to incorporate prior information into FWI to help overcome the aforementioned inversion challenges \cite{asnaashari2013regularized,zhang2019regularized,li2021deep,wang2023prior}. Various regularization techniques are designed to penalize undesirable features or promote desirable ones in the subsurface model. For example, Tikhonov \cite{golub1999tikhonov} and Total Variation (TV; \cite{strong2003edge}) regularization techniques are commonly used to, respectively, enhance smoothness and preserve sharp interfaces while reducing artifacts. In recent years, deep learning has emerged as a powerful tool for injecting prior information into FWI \cite{mousavi2022deep}. \cite{zhang2019regularized} and \cite{li2021deep} developed a deep neural network to combine information from seismic inversion and well logs to construct the prior model for regularizing the inversion process. \cite{li2023self} proposed a self-supervised pre-training and fine-tuning scheme for training the vision transformer (ViT) to learn the optimal mapping between seismic volume and well information. A generative adversarial network (GAN) and a diffusion model were also introduced to learn prior model information for regularizing the FWI process \cite{mosser2020stochastic,wang2023prior,zhang2024diffusionvel}. These regularized FWI methods help to estimate a subsurface model that not only fits the seismic observations, but also adheres to the prior knowledge of the subsurface properties. However, such methods tend to output a deterministic estimate of the subsurface model without the capability for uncertainty quantification.


Probabilistic inversion methods, based on Bayesian theory and statistical frameworks, provide a reasonable way to
quantifying uncertainty in seismic inversion results by estimating the posterior probability distribution of subsurface
models, given the observed data. This enables a more comprehensive exploration of the model space and offers a
probabilistic understanding of model parameters. These methods include the Monte Carlo (MC) sampling method \cite{robert1999monte},
variational inference \cite{blei2017variational} and many others. The MC sampling method, which uses random sampling
to estimate the posterior distribution of model parameters, has been applied in FWI \cite{biswas20172d,gebraad2020bayesian}. However, they are
computationally expensive and intractable for large-scale models due to the curse of dimensionality. An alternative is
variational inference, which approximates the posterior distribution by optimizing a simpler, parameterized distribution
to minimize the Kullback-Leibler divergence from the true posterior distribution. The variational inference
methods has been applied in FWI \cite{zhang2020variational}. Although variational inference methods are generally more efficient than traditional Monte Carlo
methods, the optimization process still requires a large number of forward simulations, resulting in high computational
cost.

The rapid rise of machine learning (ML) has brought an innovative and promising research dimension to the field of FWI. Scholars have proposed directly mapping observed seismic data to subsurface models by training neural networks on large datasets in a supervised manner \cite{yang2019deep,kazei2021mapping}.
The seismic observations need to be generated from subsurface models by numerically solving the seismic wave equation. Once the training process is complete, these approaches can directly predict subsurface models from observed seismic data without the need to solve the wave equation. However, such supervised learning techniques prevent the trained models from generalizing well to new seismic data acquired with different configurations. Besides, without a proper constraint from physical knowledge, i.e., the seismic wave equation, these methods may fail to capture complex subsurface model accurately. To address these limitations, recent research suggests combining physical knowledge with deep learning techniques in FWI \cite{mousavi2022deep,izzatullah2024physics}. \cite{wu2019parametric} and \cite{zhu2022integrating} represented the subsurface velocity model with a convolution neural network (CNN), which is then trained by optimizing the data misfit between seismic observed data and simulated data. \cite{mosser2020stochastic} applied a GAN to learn prior information representing subsurface geological features, which is then combined with a PDE-constrained inverse problem. 

Generative models, including generative adversarial networks (GANs), variational autoencoders (VAEs), normalizing flows, and diffusion models, have become indispensable tools in ML. Among these, diffusion models in particular have emerged as a powerful framework, due to their capacity to produce high-fidelity and diverse samples through stable training. They have provided a new promising way to solve inverse problems, such as image restoration and reconstruction, medical imaging, weather forecast, geophysical inversion and more \cite{song2021solving,chung2022improving,price2025probabilistic,li2024conditional,zhang2024conditional}. \cite{wang2023prior} trained a diffusion model to learn prior information of the subsurface, and subsequently the trained model are used to regularize the FWI process. \cite{zhang2024diffusionvel} integrated multiple information, including seismic data, geological knowledge, background model and well logs, in an effective way to estimate the subsurface velocity model by using a generative diffusion model, referred to as DiffusionVel. The diffusion model and FWI are closely connected through their underlying probability theory, enabling the development of an effective inversion workflow within a unified framework that combines the diffusion model and FWI.

 In this study, we propose DiffusionInv, an approach for predicting the subsurface model along with its posterior distribution from seismic data by integrating physics knowledge and prior model information. We first train the diffusion model on velocity samples that are prepared to represent our geological prior to learn the prior knowledge, which forms the samples of our prior distribution in which we hope the network will learn. Then, we fine-tune the diffusion model by using an L2 norm of data residuals between PDE-simulated data and seismic observed data as the loss function. With this FWI guidance, the model starts to learn the posterior probability distribution given seismic observations starting from the learned prior distribution. The posterior probability distribution provides valuable insights into the inversion uncertainty for PDE-constrained FWI given the observed seismic data and prior model information. Compared to regularized FWI approaches, the probabilistic framework of diffusion models enables adaptable integration of prior geological knowledge through its Bayesian inference mechanism and iterative denoising process. This mitigates the non-uniqueness problem inherent in ill-posed inverse problems, yielding geologically consistent velocity models with reduced artifacts. In contrast to purely data-driven FWI methods, DiffusionInv embeds physical constraints, i.e., seismic wave equation, directly into its generative architecture, reducing the reliance on extensive training datasets, especially seismic data, while ensuring generalization across variable acquisition geometries and subsurface heterogeneities. Finally, we analyze and test the performance of the DiffusionInv method using the Hess and Otway models.





\section{Theory and Methodology}
The proposed DiffusionInv method provides a probabilistic solution to the PDE-constrained optimization problem by integrating a pretrained diffusion model into full waveform inversion (FWI) within a Bayesian inversion framework. In the following section, we begin with setting up the optimization problem of FWI, followed by an introduction to the DiffusionInv method from the perspective of Bayesian inversion theory. Next, we explain how to learn and sample a target distribution with the diffusion model. Finally, we present the workflow and architecture of the proposed DiffusionInv method.

\subsection{The optimization problem}

FWI is a least-square data fitting procedure to estimate the optimal model \( \mathbf{m}^* \) that represents the subsurface medium for synthetic data that fits the observed ones. It can be formulated as the following optimization problem:

\begin{equation}\label{eq:1}
\mathbf{m}^* = \arg \min_\mathbf{m \in \mathcal{M}} \frac{1}{2} \| \mathbf{F}(\mathbf{m}) - \mathbf{d}_{\text{obs}} \|_2^2,
\end{equation}

where \( \mathbf{d}_{\text{obs}} \) is the observed seismic data, \( \| \cdot \|_2 \) represents the L2 norm, \(\mathbf{F}\) is a seismic wave equation operator for simulating seismic data given a subsurface model \(\mathbf{m}\) belonging to model space \(\mathcal{M}\). For simplicity and practically reasons and as often done, we use the acoustic wave equation as the governing equation for seismic wave propagation in the subsurface, which is given by the following second-order partial differential equation:
\begin{equation}\label{eq:2}
\frac{1}{v^2(\mathbf{x})} \frac{\partial^2 u(\mathbf{x},t)}{\partial t^2} - \Delta u(\mathbf{x},t) = \delta(\mathbf{x} - \mathbf{x}_s) f(t) ,
\end{equation}
where \( u(\mathbf{x},t) \) represents the seismic wavefield, \( \mathbf{x}_s \) is the source location, \( f(t) \) is the source wavelet, and \( v(\mathbf{x}) \) is the seismic wave velocity. The inverse problem is thus a PDE-constrained nonlinear optimization problem, which is iteratively solved using gradient-based optimization algorithms. Additional regularization terms can be added to the objective function in equation 1 to incorporate prior information into the inversion process.

\subsection{DiffusionInv theory}

Based on the Bayesian framework, the posterior probability distribution over the subsurface model $\mathbf{m}$ given the observed data $\mathbf{d}_{obs}$ is approximately given by:
\begin{equation}\label{eq:3}
    p(\mathbf{m}|\mathbf{d}_{obs})\propto p(\mathbf{d}_{obs}|\mathbf{m})p(\mathbf{m}),
\end{equation}
where $p(\mathbf{m})$ is the prior probability density function (PDF), capturing our prior knowledge or expectation of the subsurface model, and $p(\mathbf{d}_{obs}|\mathbf{m})$ is the likelihood, describing the probability of obtaining the observed data $\mathbf{d}_{obs}$ given a subsurface model $\mathbf{m}$ as predicted by the forward process. In equation 3, the posterior distribution provides a PDF of subsurface models given the observed data, and thus encompasses information about the uncertainty involved in subsurface inversion results. In other words, the prior distribution is updated by our observed data through the likelihood to provide us with the posterior distribution. Here, we will use the Diffusion model to store the distributions, and use the FWI mechanism to incorporate the likelihood through fine-tuning the Diffusion model. 

Monte Carlo sampling methods are often used to solve Bayesian inverse problems. However, due to the curse of dimensionality, the number of samples needed to properly sample the posterior grows exponentially with the dimensionality of inverse problem, making such methods computationally intractable for large-scale models. To overcome this limitation, we combine diffusion models with FWI to solve the inverse problem. Unlike Monte Carlo methods, which explore the model space through random sampling to build the posterior distribution, diffusion models learn to store the prior distribution; then through the FWI iterative process applied to their parameters, these Diffusion models will acquire the approximate features of the posterior distribution, in a direct manifestation of equation 3. We thus propose two steps in the DiffusionInv method for it to approximate the posterior distribution. 
\begin{itemize}
  \item First, we learn the prior distribution $p(\mathbf{m})$ by training a diffusion model on velocity model samples that are prepared based on prior knowledge of the subsurface. 
  \item Second, we fine-tune the diffusion model by using the L2 norm of data residuals between PDE-simulated data and seismic observed data as the loss function. 
\end{itemize}
This process enables the model to elegantly learn the posterior probability distribution given observed data starting from the learned prior distribution.

 Diffusion models are a family of probabilistic generative models that progressively degrade data by injecting noise and then learn to reverse this process for sample generation. Diffusion models include three predominant sub-categories: denoising diffusion probabilistic models (DDPMs), score-based generative models (SGMs), and stochastic differential equations (Score SDEs) \cite{ho2020denoising, song2020score}. In this study, we utilize DDPMs to learn the target probabilistic distributions of the velocity models used in their training. Once trained, the diffusion model is able to generate new samples from the target velocity distribution.

\subsection{Diffusion models for learning and sampling a data distribution }

 A denoising diffusion probabilistic model (DDPM) consists of a forward diffusion process and a reverse diffusion process. An illustration of the forward and reverse processes of the DDPM is shown in Figure 1. The forward process gradually adds Gaussian noise to the training sample (e.g., $\mathbf{m}_0$ ) to transform the data distribution (e.g., $q(\mathbf{m}_0)$ ) to a simple Gaussian distribution. According to the Markov chain, given a data distribution $\mathbf{m}_0 \sim q(\mathbf{m}_0)$, we can obtain the joint distribution of $\mathbf{m}_1,\mathbf{m}_2...\mathbf{m}_T$ conditioned on $\mathbf{m}_0$ by using the transition kernel $q(\mathbf{m}_t|\mathbf{m}_{t-1})$,
\begin{equation}\label{eq:4} 
\begin{aligned}
 \quad q({\mathbf{m}}_{1,\dots,T}|{\mathbf{m}}_0)&=\prod\limits_{t=1}^T q({\mathbf{m}}_t|{\mathbf{m}}_{t-1}), \text{in which}\\ 
  q(\mathbf{m}_t|\mathbf{m}_{t-1})&=\mathcal{N}(\mathbf{m}_t;\sqrt{1-\beta_t}\mathbf{m}_{t-1},\beta_t\mathbf{I}),
\end{aligned}
\end{equation} 
where the transition kernel $q({\mathbf{m}}_t|{\mathbf{m}}_{t-1})$ is designed as a Gaussian transition to slowly transform $q(\mathbf{m}_0)$ to a Gaussian distribution over T timesteps, $\mathbf{m}_t$ is the noised sample at timestep $t$, and $\beta_t$ is a predefined noise schedule that increases with $t$. The Gaussian transition kernel allows us to directly obtain the analytical form of $q(\mathbf{m}_t|\mathbf{m}_{0})$ given by
\begin{equation}\label{eq:5} 
\begin{aligned}
  q(\mathbf{m}_t|\mathbf{m}_{0})&=\mathcal{N}(\mathbf{m}_t;\sqrt{\bar\alpha_t}\mathbf{m}_{0},(1-\bar\alpha_t) \mathbf{I}),
\end{aligned}
\end{equation} 
where $\bar\alpha_t = \prod\limits_{s=1}^t (1-\beta_s)$. Given $\mathbf{m}_0$ and random noise $\boldsymbol{\epsilon}\sim\mathcal{N}(\mathbf{0},\mathbf{I})$, we can easily compute a sample of $\mathbf{m}_t$ within one step as shown below,
\begin{equation} \label{eq:6}
\mathbf{m}_t=\sqrt{\bar{\alpha}_t}\mathbf{m}_0+\sqrt{1-\bar{\alpha}_t}\boldsymbol{\epsilon}.
\end{equation}
At timestep $T$, $\mathbf{m}_T$ is assumed to be random noise sampled from a normal distribution $q({{\mathbf{m}}_T}) \approx \mathcal{N}(\mathbf{m}_T; \mathbf{0}, \mathbf{I})$. Conversely, the reverse process progressively removes noise from a random noise sample from $\mathbf{m}_T\sim\mathcal{N}(\mathbf{0},\mathbf{I})$, reconstructing a new data sample from the distribution of \( p(\mathbf{m}_0) \). We can perform this reverse process by running a parameterized Markov chain with transitions \(p_\theta(\mathbf{m}_{t-1}|\mathbf{m}_t)\). Specifically, the reverse Markov chain is defined as:
\begin{equation} \label{eq:7}
p_\theta(\mathbf{m}_{0})=p(\mathbf{m}_T)\prod\limits_{t=1}^Tp_\theta(\mathbf{m}_{t-1}|\mathbf{m}_t),
\end{equation}
where \(p(\mathbf{m}_T)\) is a predefined standard Gaussian distribution, i.e, \(p(\mathbf{m}_T) = \mathcal{N}(\mathbf{m}_T; \mathbf{0}, \mathbf{I})\), \(p_\theta(\mathbf{m}_{t-1}|\mathbf{m}_t)\) is an unknown Gaussian distribution, describing the probability transition of a data sample from the current state $\mathbf{m}_{t}$ to the previous state $\mathbf{m}_{t-1}$. We can approximate the transition distribution $p_\theta(\mathbf{m}_{t-1} | \mathbf{m}_t)$ by training neural networks to learn its mean \(\boldsymbol\mu_\theta(\mathbf{m}_t, t)\) and covariance \(\Sigma_\theta\). Specifically, $p_\theta(\mathbf{m}_{t-1} | \mathbf{m}_t)$ takes the form of
\begin{equation} \label{eq:8}
p_\theta(\mathbf{m}_{t-1} | \mathbf{m}_t) = \mathcal{N}(\mathbf{m}_{t-1}; \boldsymbol\mu_\theta(\mathbf{m}_t, t), \Sigma_\theta(\mathbf{m}_t, t)).
\end{equation}

Then, we can train the neural network by minimizing the variational lower bound (VLB) on the negative log likelihood, given by:
\begin{equation} \label{eq:vlb}
 L_{\text{VLB}} = -\log p_\theta(\mathbf{m}_0 |\mathbf{m}_1) + \sum_{t>1} \text{KL} \left( q(\mathbf{m}_{t-1} | \mathbf{m}_{t},\mathbf{m}_{0}) \| p_\theta(\mathbf{m}_{t-1} | \mathbf{m}_{t})\right) \\
 +  \text{KL} \left( q(\mathbf{m}_{T} | \mathbf{m}_{0},\mathbf{m}_{0}) \| p(\mathbf{m}_{T} \right),
\end{equation}
where $\text{KL}$ stands for the Kullback-Leibler divergence for computing the distance between two probability distributions. We can see that the VLB is expressed as a sum of three independent terms. These three terms denote the reconstruction term, the consistency term and the prior matching term, respectively. The reconstruction and prior matching terms do not depend on the model parameters of the neural network, thus they can be ignored. The consistency term reflects the objective of the training process: optimizing the neural network to ensure that the learned reverse process at each time step $t$, $p_\theta(\mathbf{m}_{t-1} | \mathbf{m}_{t})$, closely match the true distribution of the forward process, $q(\mathbf{m}_{t-1} | \mathbf{m}_{t},\mathbf{m}_{0})$, which is conditioned on the original sample. The posterior distribution $q(\mathbf{m}_{t-1} | \mathbf{m}_{t},\mathbf{m}_{0})$ at each time step is a Gaussian distribution with the mean $\boldsymbol{\mu}_{q}(\mathbf{m}_{t},\boldsymbol{\epsilon})$ and a constant covariance determined by the predefined noise schedule. The mean $\boldsymbol{\mu}_{q}$ is explicitly formulated as
\begin{equation} \label{eq:mut}
\boldsymbol{\mu}_{q}(\mathbf{m}_{t}, \boldsymbol{\epsilon}_{t}) = \frac{1}{\sqrt{\alpha_t}} \left( \mathbf{m}_{t} - \frac{1-\alpha_t}{\sqrt{1 - \bar{\alpha}_t}} \boldsymbol{\epsilon} \right).
\end{equation}
Minimizing the KL divergence between the two Gaussian distributions, $p_\theta(\mathbf{m}_{t-1} | \mathbf{m}_{t})$ and $q(\mathbf{m}_{t-1} | \mathbf{m}_{t},\mathbf{m}_{0})$, can be further simplified as minimizing the difference between their means. However, the mean of $p_\theta(\mathbf{m}_{t-1} | \mathbf{m}_{t})$, denoted as $\boldsymbol\mu_\theta(\mathbf{m}_t, t)$, is not directly predicted by the neural network. Instead, the neural network is trained to predict the noise $\boldsymbol{\epsilon}$, and since $\mathbf{m}_{t}$ is already known in the reverse process, this allows the mean to be computed accordingly. Thus, the loss function used for optimizing the network parameters $\theta$ is written by
\begin{equation}\label{eq:l2}
    \operatorname*{min}_{\theta} \mathcal{L}(\theta) := \operatorname*{min}_{\theta} \mathbb{E}_{t,\mathbf{m}_t,\boldsymbol{\epsilon}}\Big[\left\|\boldsymbol{\epsilon}-\boldsymbol{\epsilon}_{\theta}(\mathbf{m}_{t},t)\right\|^2\Big].
\end{equation}
Given the prepared training dataset, the number of time steps T, and the predefined noise schedule, we can perform the training process as follows. First, we randomly sample data samples from the training dataset, as well as sample Gaussian noise $\boldsymbol{\epsilon} \sim \mathcal{N}(\mathbf{0},\mathbf{I})$ and a time step $t$. Noisy samples $\mathbf{m}_{t}$ can then be computed using equation 6. At last, we can optimize the parameters of the neural network, which receives $\mathbf{m}_{t}$ and $t$ as input, to minimize the loss function shown in equation 11.

Once the training is complete, we generate samples from the learned data distribution by sampling from the base Gaussian distribution and then applying the reverse diffusion process to transform these samples into clean samples from the original training set distribution (the velocity models we trained with). In the sampling process of DDPM, the Gaussian noise is iteratively predicted by the neural network and removed from the noisy sample over $T$ time steps. A slight modification given by a Denoising Diffusion Implicit Models (DDIM) framework \cite{song2020denoising} enhances the sampling efficiency while retaining the training framework. The deterministic sampling strategy of the DDIM allows for high-quality generation within significantly fewer steps. Thus, we employ the reverse process of DDIM, given by the following update equation, to generate new samples from the learned target distribution:
\begin{equation}\label{eq:sam}
\begin{aligned}
\mathbf{m}_{0_{\theta}}(\tau)=\frac{\mathbf{m}_{\tau}-\sqrt{1-\alpha_{\tau}}\boldsymbol{\epsilon}_{\theta}(\mathbf{m}_{\tau},\tau)}{\sqrt{\alpha_{\tau}}},
\end{aligned}
\end{equation}.
\begin{equation}\label{eq:sam1}
\begin{aligned}
\mathbf{m}_{\tau-1}=\sqrt{\alpha_{\tau-1}}\mathbf{m}_{0_{\theta}}(\tau)+&\sqrt{1-\alpha_{\tau-1}-\sigma_{\tau}^{2}}\boldsymbol{\epsilon}_{\theta}(\mathbf{m}_{\tau},\tau)+\sigma_{\tau}\boldsymbol{\epsilon},
\end{aligned}
\end{equation}
where $\tau$ is the timestep sampled from a  sub-sequence of original time sequence $\left[0,1, \dots T\right]$,  and $\sigma_{\tau}=\sqrt{(1-\alpha_{\tau-1})/(1-\alpha_{\tau})}\sqrt{(1-\alpha_{\tau}/\alpha_{\tau-1})}$, $\mathbf{m}_{0_{\theta}}$ is the estimated clean sample.


\subsection{DiffusionInv: prior-information pretraining and physics-informed fine-tuning}

The proposed DiffusionInv method is trained in two stages: (1) prior-information pretraining hopefully with geologically plausible models and (2) physics-informed fine-tuning, ensuring that the generated velocity models are physically consistent with the observed seismic data. A graphical illustration of the DiffusionInv method is shown in Figure 2. A latent diffusion model \cite{rombach2022high} is used to improve the efficiency of diffusion-based generative models by performing the diffusion process in a lower-dimensional latent space rather than the original high-dimensional data space (velocity models). The latent diffusion model first use a pretrained autoencoder to encode the input velocity models into a latent representation. Both the forward diffusion and reverse denoising processes are performed in the latent space. After denoising, the latent representation is decoded back into the original data space using the pretrained decoder, reconstructing the high-resolution velocity models. U-Net is a widely used backbone in diffusion models, particularly for predicting the Gaussian noise in reverse denoising process. Figure 3 shows the U-shaped network architecture of our used U-Net, which has shown good performance in our previous work \cite{zhang2024conditional, li2024conditional}. The architecture mainly consists of an encoder, a decoder, and a bottleneck connecting them. The U-Net takes the noisy latent representation and time-step embedding as inputs and outputs the predicted noise. The ResNet blocks in the encoder extract hierarchical features from the input latent, while those in the decoder upsample the feature maps to restore the original input size. The bottleneck integrates ResNet blocks and self-attention mechanisms to capture long-range dependencies.

When prior knowledge of the subsurface - such as geological structures, velocity distribution, and well logs - is available, velocity model samples can be built to represent our empirical expectation of the subsurface velocity. These velocity model samples are then used to train the diffusion model to learn a prior distribution $p(\mathbf{m})$ in equation 3. In the training process, Gaussian noise is gradually added to the latent representation of these velocity models through forward diffusion process, and the U-Net learns to reverse this process by denoising to capture the prior distribution $p(\mathbf{m})$ effectively. Once trained, the diffusion model is able to efficiently generate velocity samples that follow the learned distribution. The learned prior distribution can then provide an effective regularization for solving the seismic inverse problem by using the observed data to guide the generation process. Here, we apply the approximate Bayesian formula (equation 3) by using the observed data to refine the representation of the prior distribution given by the trained Diffusion model to approximate the posterior distribution $p(\mathbf{m}|\mathbf{d}_{obs})$ of the velocity models.

During the fine-tuning process, the pretrained network is refined by incorporating the seismic observations and the physical constraint given by the seismic wave equation. Specifically, the L2 norm of the data residuals between PDE-simulated seismic data $\mathbf{d}_{syn}=\mathbf{F}(\mathbf{m})$ and observed data $\mathbf{d}_{\text{obs}}$ is used as the loss function to update model parameters. This physics-constrained loss is formulated as
\begin{equation}\label{loss1}
    \operatorname*{min}_{\theta} \mathcal{L}(\theta) := \operatorname*{min}_{\theta} \mathbb{E}_{\mathbf{m}_T} \left\| \mathbf{F}(p_\theta(\mathbf{m}_{0})) - \mathbf{d}_{\text{obs}} \right\|^{2},
\end{equation}
where $\mathbf{F}$ is the wave equation operator that simulates synthetic data based on the generated velocity models. We generate velocity models by sampling from the learned prior distribution and obtain the corresponding synthetic data by using DeepWave \cite{richardson_alan_2023}, and then measure the loss between these synthetic data and the observed ones. The residuals are then back-propagated through automatic differentiation to update the parameters of the diffusion model. After this optimization, we anticipate that the diffusion model adjusted the reverse process (given by the network) to map from a Gaussian distribution $p(\mathbf{m}_T)$ to an approximation of the posterior distribution $p(\mathbf{m}|\mathbf{d}_{\text{obs}})$, incorporating both prior model information and seismic data. However, directly re-training the diffusion model can disrupt the prior information encoded in the model, leading to instability in the learned velocity distribution.

To prevent degradation of prior information while fine-tuning for posterior learning, we adopt parameter-efficient fine-tuning techniques. Specifically, we update the U-Net decoder while freezing the other model parameters, thereby preserving the prior knowledge while adapting the model to seismic data. A low learning rate is used to ensure smooth adaptation without significant deviation from the original prior information constraints. The modified loss function for the PDE-constrained fine-tuning is written as
\begin{equation}\label{loss2}
   \operatorname*{min}_{\theta_2} \mathcal{L}(\theta_1,\theta_2) := \operatorname*{min}_{\theta_2} \mathbb{E}_{\mathbf{m}_T} \left\lVert \mathbf{F}(p_{\theta_1,\theta_2}(\mathbf{m}_{0:T})) - \mathbf{d}_{\text{obs}} \right\rVert^{2}   \quad \text{(with \(\theta_1\) fixed)}
\end{equation}
where $\theta_{1}$ and $\theta_2$ stands for the frozen and activate parameters, respectively. This fine-tuning process optimizes the physics consistency with seismic observations while ensuring consistency with prior model information, thereby improving inversion accuracy and robustness. 

  \begin{figure}[htpb!]
    \centering
    \includegraphics[width=0.98\columnwidth]{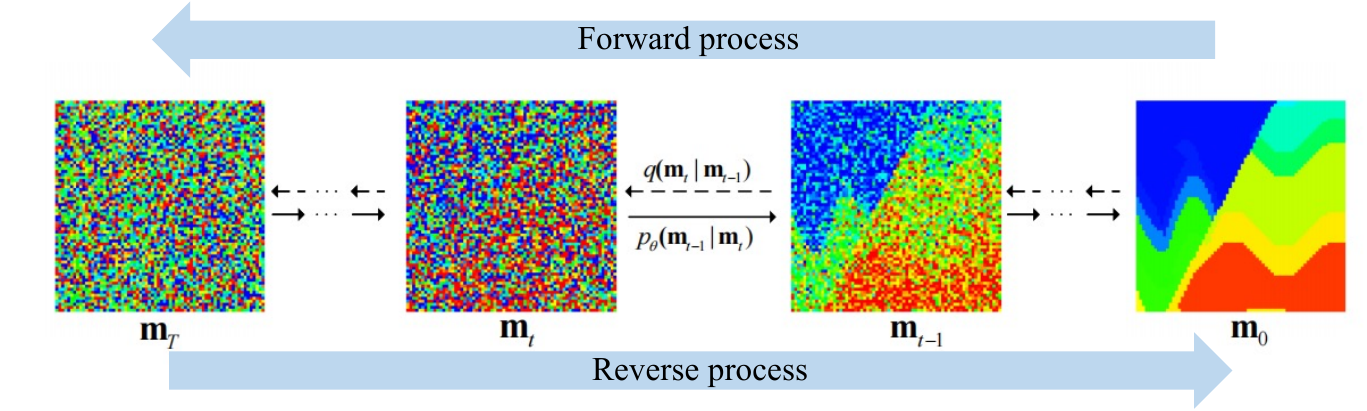}
    \caption{The forward and reverse processes of the denoising diffusion probabilistic model (DDPM).
}
    \label{fig:diffusionmodel}
\end{figure}

\begin{figure}[htpb!]
    \centering
    \includegraphics[width=0.98\columnwidth]{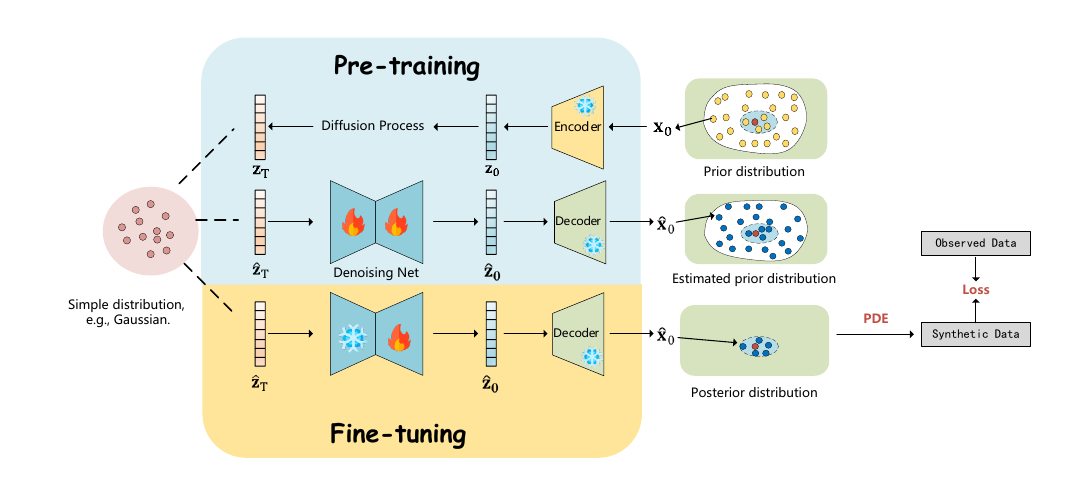}
    \caption{The architecture of the proposed DiffusionInv method. A latent diffusion model is used to improve the generation efficiency by using a pretrained autoencoder to encode the input into a lower-dimensional latent representation and performing the diffusion process in this latent space. The encoder maps input velocity model samples into a latent space, capturing their low-dimensional representations, while the decoder reconstructs the velocity model from the latent space. The pretraining process (blue box) consists of a forward diffusion process and a reverse denoising process, where the diffusion model is trained on velocity model samples (yellow circles) prepared based on prior model knowledge to learn the prior distribution. In the fine-tuning stage (yellow box), the decoder layers of the pretrained diffusion model are further updated by minimizing the misfit between observed and PDE-simulated data for learning the posterior distribution of the velocity model given the observed data. 
}
    \label{fig:workflow}
\end{figure}

\begin{figure}[htpb!]
    \centering
    \includegraphics[width=0.7\columnwidth]{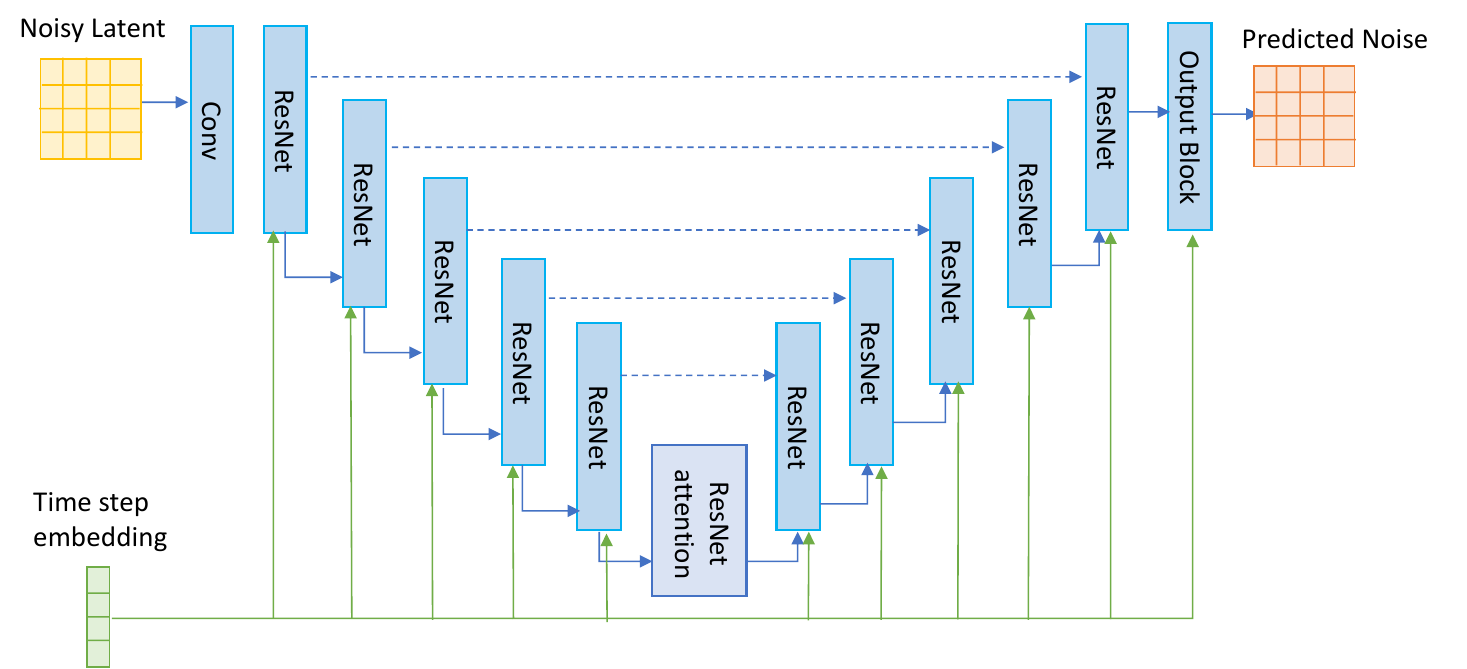}
    \caption{The architecture of our used U-Net. 
}
    \label{fig:unet}
\end{figure}

\section{Experiments}
In this section, we test our proposed DiffusionInv method on two synthetic datasets to demonstrate how the proposed DiffusionInv solves the FWI problem with improved accuracy, enhanced resolution, and estimation of inversion uncertainty.

\begin{figure}[htpb!]
    \centering
    \includegraphics[width=0.90\columnwidth]{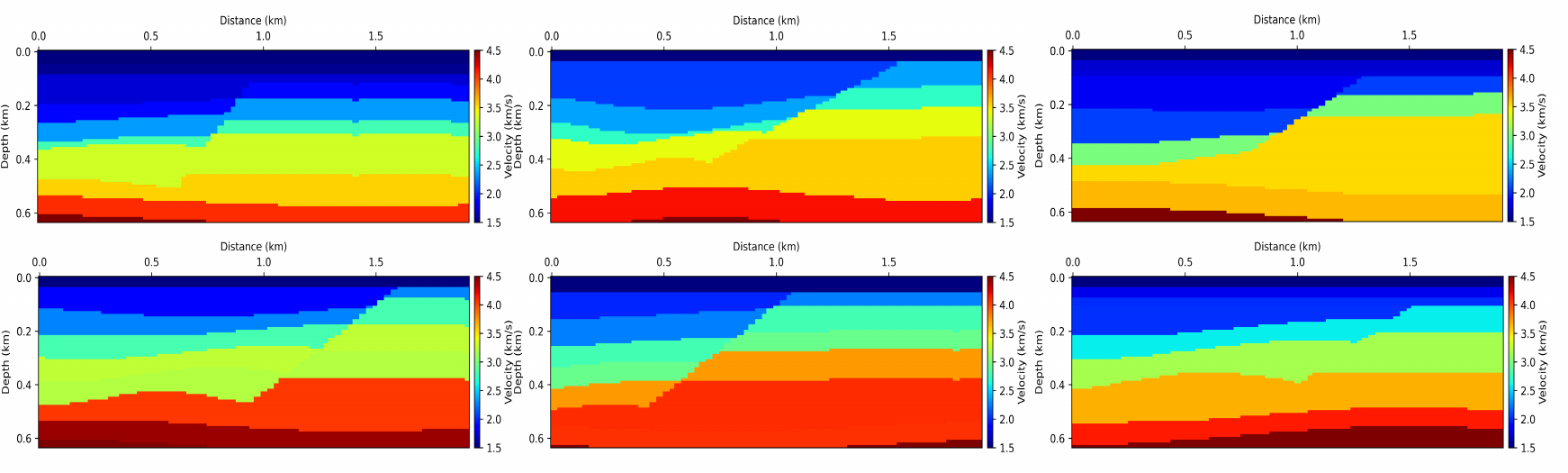}
    \caption{The Hess model: six example samples of the prepared velocity models for training.}
    \label{fig:hess_sam_true}
\end{figure}

\begin{figure*}[htpb!]
    \centering
    \subfloat[]{\includegraphics[width=0.5\columnwidth]{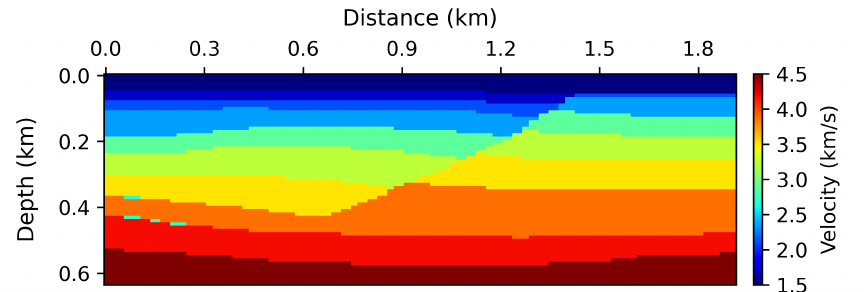}\label{fig:true}}
    \subfloat[]{\includegraphics[width=0.5\columnwidth]{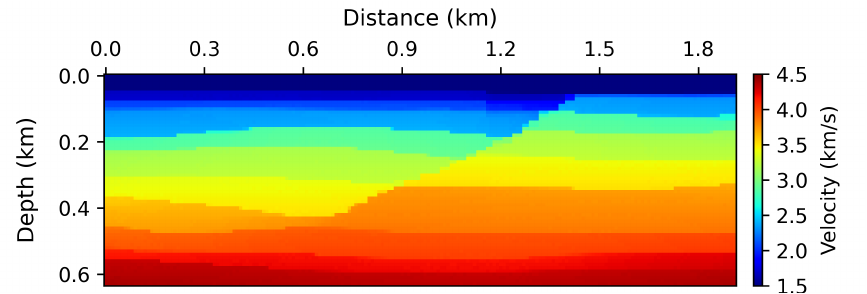}\label{fig:diffusioninv}}\\
    \subfloat[]{\includegraphics[width=0.5\columnwidth]{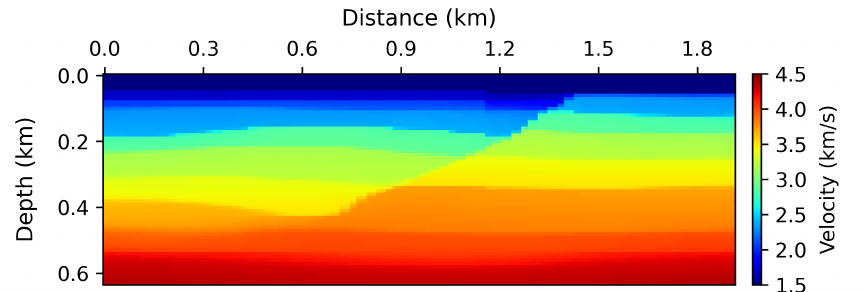}\label{fig:mean}}
    \subfloat[]{\includegraphics[width=0.5\columnwidth]{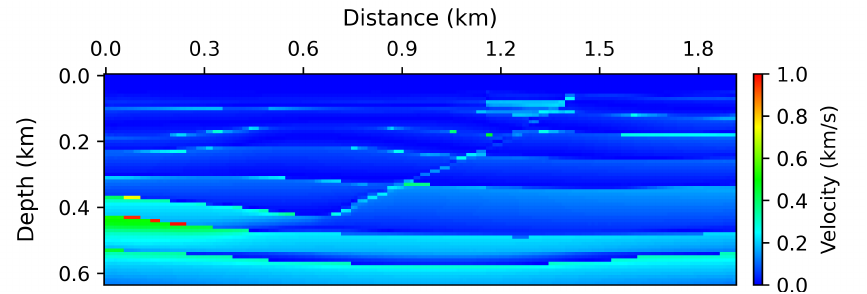}\label{fig:meanerr}}\\
    \subfloat[]{\includegraphics[width=0.5\columnwidth]{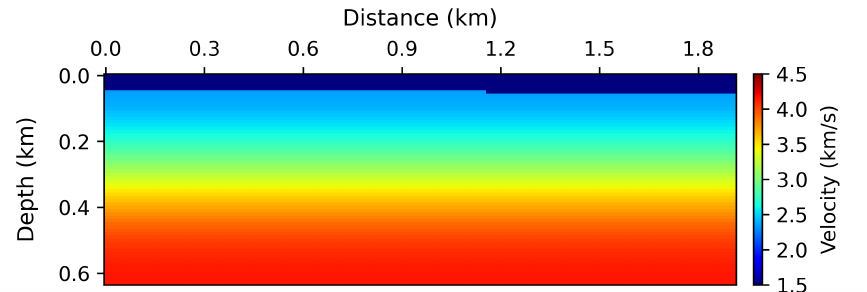}\label{fig:initial}}
    \subfloat[]{\includegraphics[width=0.5\columnwidth]{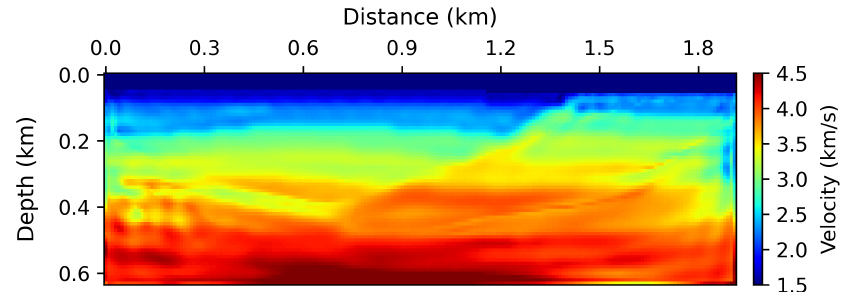}\label{fig:fwi}}\\

    \caption{The Hess model example: (a) true velocity, (b) one example velocity sampling from the approximate posterior distribution, (c) the mean velocity of the approximate posterior distribution, (d) the velocity error between the mean velocity and the true one, (e) the initial velocity for conventional FWI, and (f) the conventional FWI result.}
    \label{fig:Invvel}
\end{figure*}

\begin{figure*}[htpb!]
    \centering
    \subfloat[]{\includegraphics[width=0.5\columnwidth]{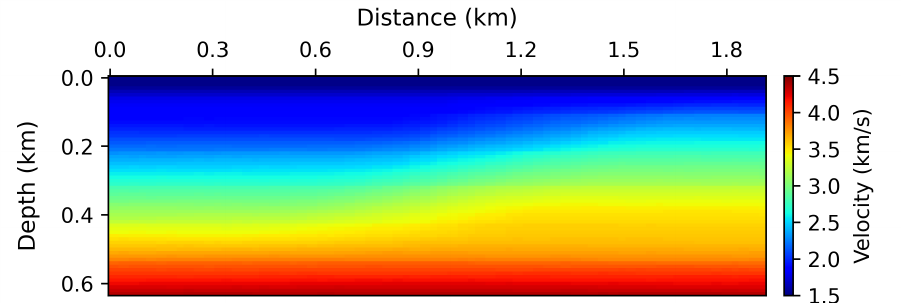}\label{fig:true}}
    \subfloat[]{\includegraphics[width=0.5\columnwidth]{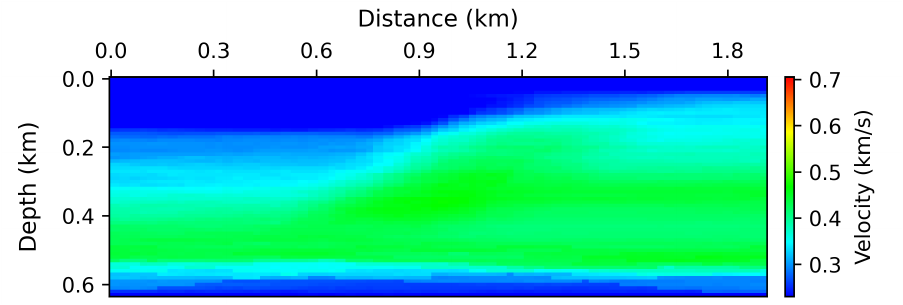}\label{fig:diffusioninv}}\\
    \subfloat[]{\includegraphics[width=0.5\columnwidth]{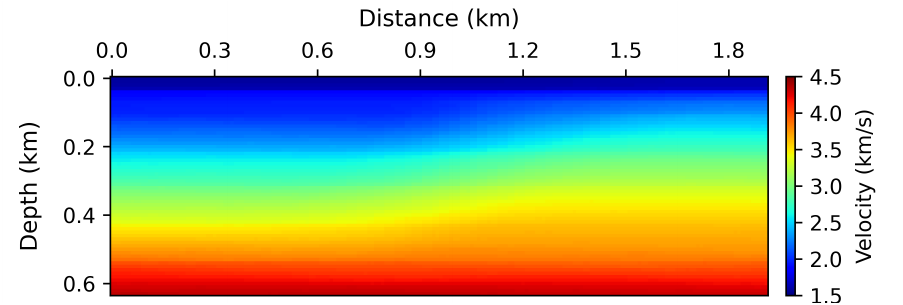}\label{fig:initial}}
    \subfloat[]{\includegraphics[width=0.5\columnwidth]{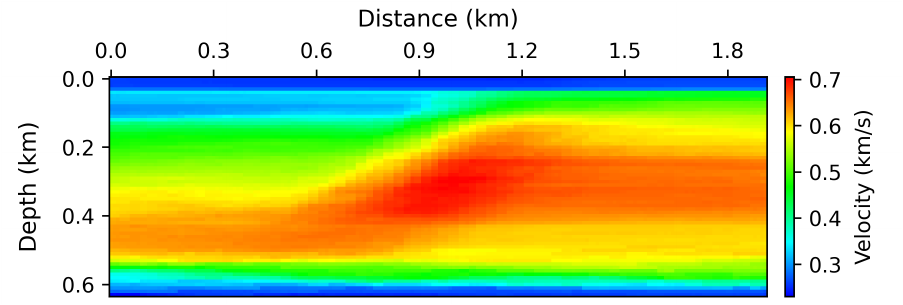}\label{fig:fwi}}\\
    
    \subfloat[]{\includegraphics[width=0.5\columnwidth]{Fig/hess_posterior_mean1.pdf}\label{fig:mean}}
    \subfloat[]{\includegraphics[width=0.5\columnwidth]{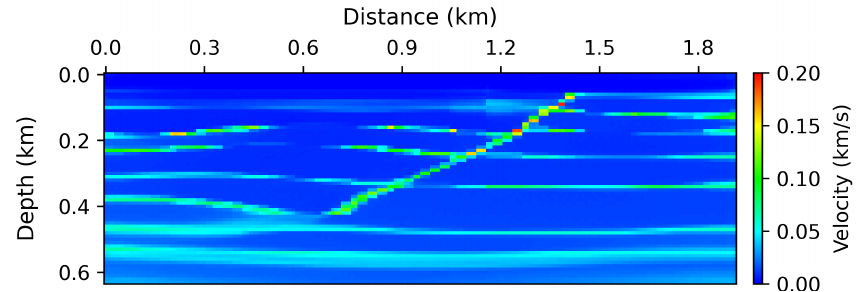}}\label{fig:std}

    \caption{The Hess model example: (a)(b) the mean and standard deviation of the underlying distribution of the training data, (c)(d) the mean and standard deviation of the learned prior distribution after pretraining, (e)(f) the mean and standard deviation of the approximate posterior distribution after finetuning.}
    \label{fig:Invvel}
\end{figure*}


\subsection{Hess model example}
To validate the effectiveness of the proposed DiffusionInv method, we first apply it on a portion of the Hess model as shown in Figure 5a. This model comprises a layered sedimentary sequence intersected by a normal fault with a dip angle of about $25^{\circ}$. We use a sparse seismic array with a total of 5 shots located at a 10 m depth to generate seismic data. A Ricker wavelet with a dominant frequency of 10 Hz is used as the source wavelet to simulate wave propagation. 

To incorporate prior model information into the inversion, we first prepare a training dataset of 4,500 velocity models that represent plausible sedimentary layer and fault structures. These models should represent our expectations of what the solution should look like, with features like faults, layers, and possibly a general increase of velocity with depth, maybe extracted from a well log. The Diffusion model is not expected to store these velocity models, but to store the features of their distribution, so it can generate random models with similar features. Figure 4 shows six representative velocity models from the training set. \textcolor{black}{We can see that the number and thickness of the layers and the dip of fault cover reasonable variations.} So, the diffusion model is subsequently pretrained on these velocity models to learn the multi-scale velocity patterns of the subsurface, including the background velocity trend, the velocity continuity in sedimentary layers, and the fault-associated velocity discontinuity. In this pre-training stage, we used the Adam optimizer, a learning rate of $5 \times 10^{-3}$ and 259 epochs. Once the prior distribution is learned, we fine-tune the U-Net decoder using a learning rate $7 \times 10^{-6}$ for 2000 iterations to minimize the L2 norm of the data residuals between the simulated and observed seismic data to approximate the posterior distribution of the subsurface velocity given observed seismic data. An example velocity model obtained by sampling from the fine-tuned diffusion model is shown in Figure 5b. We can see that the expected features represented by the prior models, such as the layers, the fault structure and the velocity trend, are preserved well in this generated velocity model sample, meanwhile this velocity looks close to the true model due to the physics constraint of observed data. To analyze the learned posterior distribution, we generate 500 random velocity model samples from the fine-tuned diffusion model and compute their mean velocity as an approximate of the posterior distribution's expected velocity model. The mean velocity is shown in Figure 5c. Figure 5d shows its velocity error map, which is the difference between the true model and the mean velocity shown in Figure 5c. We can see that the proposed DiffusionInv method successfully reconstructed the fine-scale velocity discontinuities with high resolution including the sharp velocity contrasts near the fault while maintaining continuity in the sedimentary layers. However, the inversion resolution and accuracy decrease in deeper regions due to weaker seismic illumination. Additionally, the low-velocity anomaly at about 0.4 km depth on the left are not recovered well, though some evidence of it can be seen, by the lower average velocity in that area. This can be attributed to the fact that such velocity structures were not represented well (very low probability) in our prior model information, but yet the data was pushing for it, obviously at a low resolution. For comparison, we perform conventional FWI using the velocity model shown in Figure 5e as an initial model. The conventional FWI inverted velocity model is shown in Figure 5f. We can see that the conventional FWI recovers the large-scale velocity structures but is affected by some inversion artifacts and struggles to resolve the fine-scale velocity discontunities at the layer interfaces and fault due to the limited frequency band of the seismic data.

To demonstrate how the probability distribution of velocity models evolved throughout the DiffusionInv pipeline, we quantify three key distributions: the underlying distribution of the training velocity models (before pretraining), the learned prior distribution after pretraining, and the approximate posterior distribution after fine-tuning. We first compute the mean and standard deviation of the velocity models used for pretraining the diffusion model to represent the underlying prior distribution. Figures 6a and 6b show such mean and standard deviation, respectively. The mean velocity shows mainly the overall trend of the velocity models, whereas the standard deviation highlights potential velocity variations. After pretraining, we can randomly generate/sample 500 velocity models from the pretrained diffusion model and obtain their mean and standard deviation (shown in Figures 6c and 6d), representing the learned prior distribution after pretraining. The mean velocity of the learned prior distribution closely aligns with that of the training velocity models, whereas the standard deviation exhibits some degree of discrepancy. The discrepancy of standard deviation reflects the diffusion model distinct learned interpretation and its low manifold representation of the training set. In a similar manner, we present in Figures 6e and 6f the mean and standard deviation of the approximate posterior distribution after finetuning. The results demonstrate that the fine-tuned diffusion model effectively learns the key characteristics of the target posterior distribution, with the mean closely resembling the true velocity (Figure 5a). It is important to note that the standard deviation quantifies the spread or uncertainty around the mean velocities. The estimated velocities in the shallow layers generally exhibit a smaller standard deviation, indicating that the posterior distribution is tightly clustered around the mean, which suggests a high level of confidence (low uncertainty) in the estimated velocities. However, the estimated velocities at fault boundaries, layer interfaces, and in deeper layers show a higher standard deviation, indicating greater uncertainty in these regions due to the weaker constraint adhered by the observed data. Furthermore, the strong spatial correlation between the posterior standard deviation and the velocity error map (Figure 5d) demonstrates its reliability for quantifying the inversion uncertainty. Thus, the DiffusionInv approach effectively obtains a reliable approximate of the posterior distribution of the subsurface velocity.

To further validate the inversion result, we show comparisons of seismic data and velocity profiles. Figure 7 shows the comparison of the simulated shot gather from the mean velocity model (left) with the observed shot gather (right). The close match between the two gathers ensures that the generated velocity model is physically consistent with the recorded seismic data. A direct comparison of the velocity profiles at locations 0.40 km and 1.60 km are shown in Figures 8a and 8b, respectively. The velocity profiles of the DiffusionInv inversion result show good consistency with the true velocity profiles, capturing the fine-scale velocity variations that are not well recovered by conventional FWI. The slight deviation of the velocities in the deep is acceptable considering their uncertainty. Overall, the DiffusionInv approach enhances the inversion resolution while providing a reliable uncertainty quantification.

\subsubsection{Influence of limited acquisition}

To further test the robustness of the DiffusionInv method, we conduct an additional test using an extremely limited acquisition setup, where only one single shot is placed on the left side of the model. This setup presents a highly underdetermined inversion problem, as the lack of seismic coverage makes it challenging to constrain subsurface structures, especially for the right part. The DiffusionInv inverted velocity for this scenario is shown in Figure 9a. Figure 9b shows the velocity error between the inverted velocity and the true velocity model. We can see that the DiffusionInv method manages to recover the layered velocities on the left side of the model with high resolution, as these velocity structures are reasonably constrained by seismic data. However, the accuracy of the inverted velocity on the right side is obviously reduced because of the lack of seismic constraints. Correspondingly, the obtained standard deviation, shown in Figure 9c, reveals increased uncertainty on the right part with poor seismic illumination. These test results demonstrate that the DiffusionInv method is effective in integrating prior model information and physical constraint of observed seismic data, leading to improved velocity reconstruction and reasonable uncertainty quantification.

\begin{figure*}[htpb!]
    \centering
    \includegraphics[width=0.5\columnwidth]{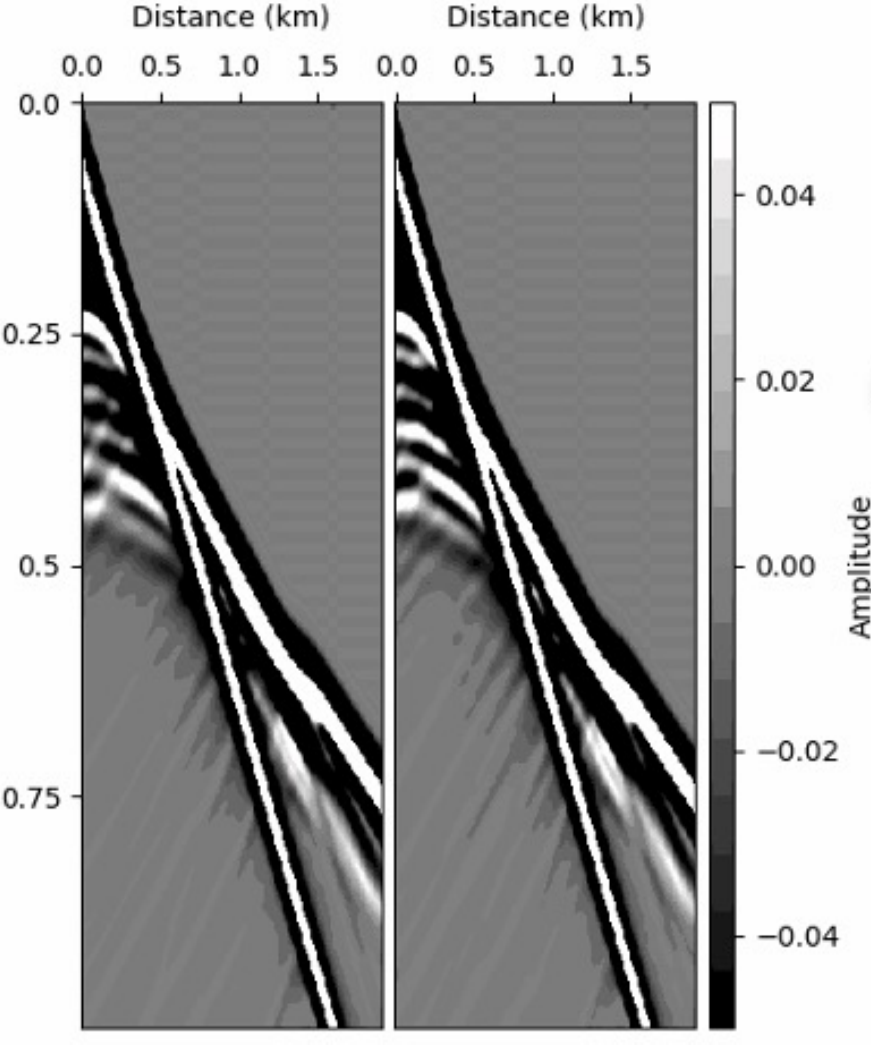}\label{fig:initial}\\
    \caption{Comparison between the simulated data from the DiffusionInv generated velocity model (left) and the observed data (right).
}
    \label{fig:Invvel}
\end{figure*}

\begin{figure*}[htpb!]
    \centering
    \subfloat[]{\includegraphics[width=0.75\columnwidth]{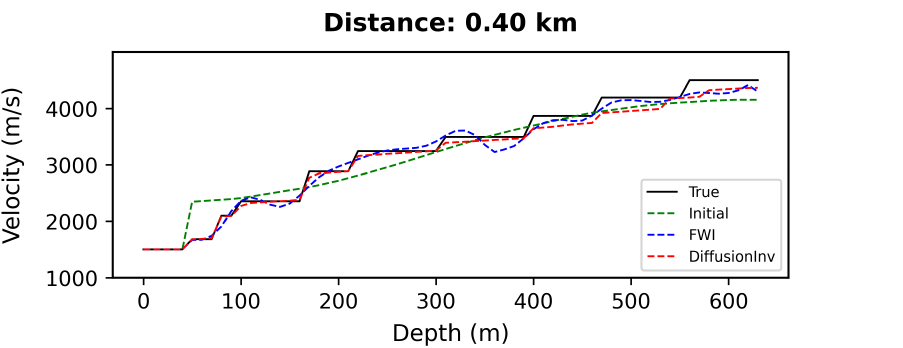}\label{fig:velpro_0d4}}\\
    \subfloat[]{\includegraphics[width=0.75\columnwidth]{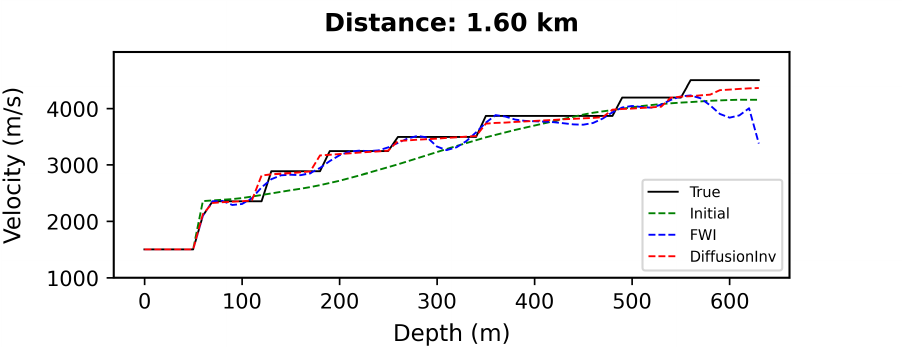}\label{fig:velpro_1d6}}\\
    \caption{The Hess model example: the velocity profiles at distance (a) 0.4 km and (b) 1.60 km.}
    \label{fig:velpro}
\end{figure*}

\begin{figure*}[htpb!]
    \centering
    \subfloat[]{\includegraphics[width=0.5\columnwidth]{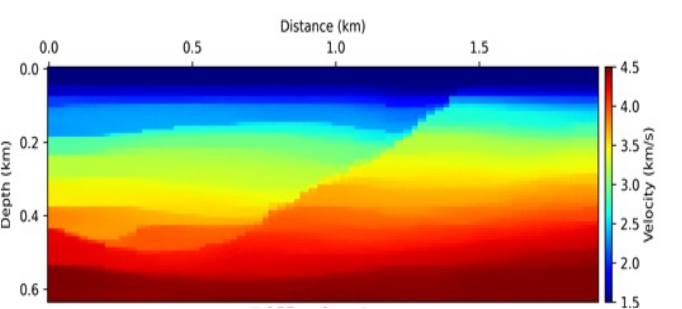}\label{fig:diffusioninv}}\\
    \subfloat[]{\includegraphics[width=0.5\columnwidth]{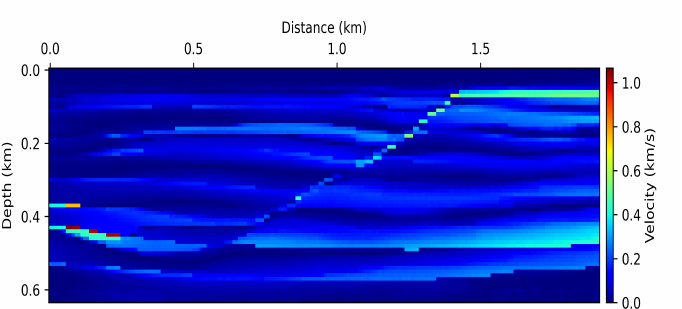}\label{fig:fwi}}
    \subfloat[]{\includegraphics[width=0.46\columnwidth]{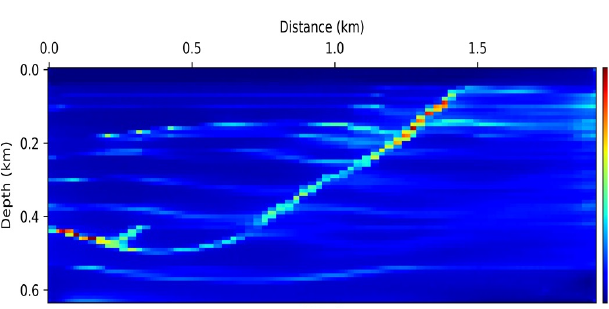}\label{fig:initial}}

    \caption{The Hess model example with very limited acquisition system (only one shot on the left): (a) the mean velocity of the approximate posterior distribution, (b) the velocity error between the mean velocity and the true model, and (c) the standard deviation of the approximate posterior distribution.}
    \label{fig:Invvel}
\end{figure*}

\subsection{Otway model example}

We then evaluate the proposed DiffusionInv method using the Otway model, as shown in Figure 13a. The Otway model represents a complex geological setting composed of numerous thin layers, making it particularly challenging for conventional FWI to recover these thin layers with high accuracy because of the limited frequency bandwidth of seismic data. For the seismic acquisition setup, we employ 10 evenly spaced sources positioned at a depth of 5 m. The seismic data for each shot is recorded by 576 receivers, which are uniformly distributed across the acquisition surface. A Ricker wavelet with a main frequency of 15 Hz is used as the source wavelet to excite seismic wave propagation through the subsurface.

To pretrain the diffusion model to learn the prior model distribution, we first prepare a training dataset of 4,500 velocity models by randomly incorporating additional perturbations into the velocity profiles from the background velocity while integrating subsurface structures inferred from imaging result. Figure 10 shows nine samples from these velocity models. We then pretrain the diffusion model on these prepared velocity samples using the Adam optimizer, allowing it to effectively capture the underlying prior distribution of velocity models. The number of training epochs is 289. To quantify and visualize the prior distribution $p(\mathbf{m})$, we compute the mean and standard deviation ($\sigma$) at two representative locations (distance: 0.55 km and 1.80 km) across all 4,500 velocity model samples and plot in Figure 11 the mean velocity curves with a shaded region representing $\pm\sigma$ to represent the distribution of the velocity profiles. The mean and the $\pm\sigma$ range of this prior distribution are represented by the blue line and the blue shaded region, respectively. Correspondingly, the learned prior distribution can be visualized by computing the mean and standard deviation for the generated velocity models from the pretrained diffusion model. The learned prior distribution is shown by the black line for the mean and the green shaded region for the $\pm\sigma$ region. We can see that the learned prior distribution closely matches the original prior distribution of the subsurface velocity model. This demonstrates that the pretrained diffusion model effectively captures the prior distribution by training on these prepared velocity model samples. Once the prior distribution has been learned, we further fine-tune the U-Net decoder by incorporating seismic observations and physical constraints. The optimization process runs for 2,000 iterations with a learning rate of $7 \times 10^{-6}$, during which the model progressively adjusts the network to better align with the observed seismic data. Figure 12 shows the convergence curve of the loss function. We can see that the loss steadily decreases to a low value, which indicates that the inverted velocity model match well with the seismic data. An example velocity model sampled from the fine-tuned diffusion model, as shown in Figure 13b, exhibits rich thin layers, capturing key structural features. The mean velocity obtained by using 500 randomly sampled velocity models is shown in Figure 13c, representing the most probable velocity model estimated by the proposed DiffusionInv method. Figure 13d shows the velocity error between the mean velocity model and the true one. We can see that the mean velocity model closely resembles the true model (Figure 13a), demonstrating that the DiffusionInv approach reconstructs the thin layers in the subsurface with high resolution, accurately capturing the fine-scale geological features that are often difficult to recover using conventional FWI approaches. 


We then compare the standard deviation of the learned prior distribution after pretraining with that of the approximate posterior distribution after fine-tuning to evaluate the impact of prior model information and seismic data constraints on the probability distribution of velocity models. Specifically, we calculate the standard deviation across 500 randomly sampled velocity models from both the pretrained and fine-tuned diffusion models, as illustrated in Figures 14a and 14b, respectively. The high standard deviation of the prior distribution (Figure 14a) indicates that the pretrained diffusion model can generate the velocity models with more variations, allowing for a sufficiently expansive search space, thus enhancing the likelihood of identifying the most plausible velocity model constrained by seismic data. The standard deviation of the approximate posterior distribution is notably reduced to a lower value due to the constraints of the seismic data, particularly in the shallow part. However, the standard deviation increases with depth because the deeper layers with weaker seismic illumination, are less constrained by seismic data.

For comparison, we perform conventional FWI using the same acquisition geometry. Figure 13e shows the initial velocity model for conventional FWI, and the resulting velocity model of the conventional FWI is shown in Figure 13f. As expected, the resolution of the FWI result is constrained by the frequency bandwidth of the observed seismic data, leading to lower resolution and difficulty in capturing fine-scale structures. Figure 15 shows a comparison between the observed and simulated shot gathers for the DiffusionInv resulting mean velocity and conventional FWI result. The shot gather generated from the DiffusionInv result exhibits good consistency with the observed seismic data, but the data misfits is still a bit higher than that of conventional FWI because the incorporated prior information may not adequately align with the actual subsurface structures. We then compare the velocity profiles at 0.55 km and 2.30 km in Figure 16. We can see that the DiffusionInv approach successfully reconstructs the thin layers with high accuracy and sharp interfaces, whereas conventional FWI estimates a smoother model that primarily captures large-scale velocity trends. This demonstrates that the prior model information and seismic physical constraint are effectively integrated in the DiffusionInv approach, thus enhancing the inversion resolution and improving the characterization of complex geological structures.

\begin{figure}[htpb!]
    \centering
    \includegraphics[width=0.90\columnwidth]{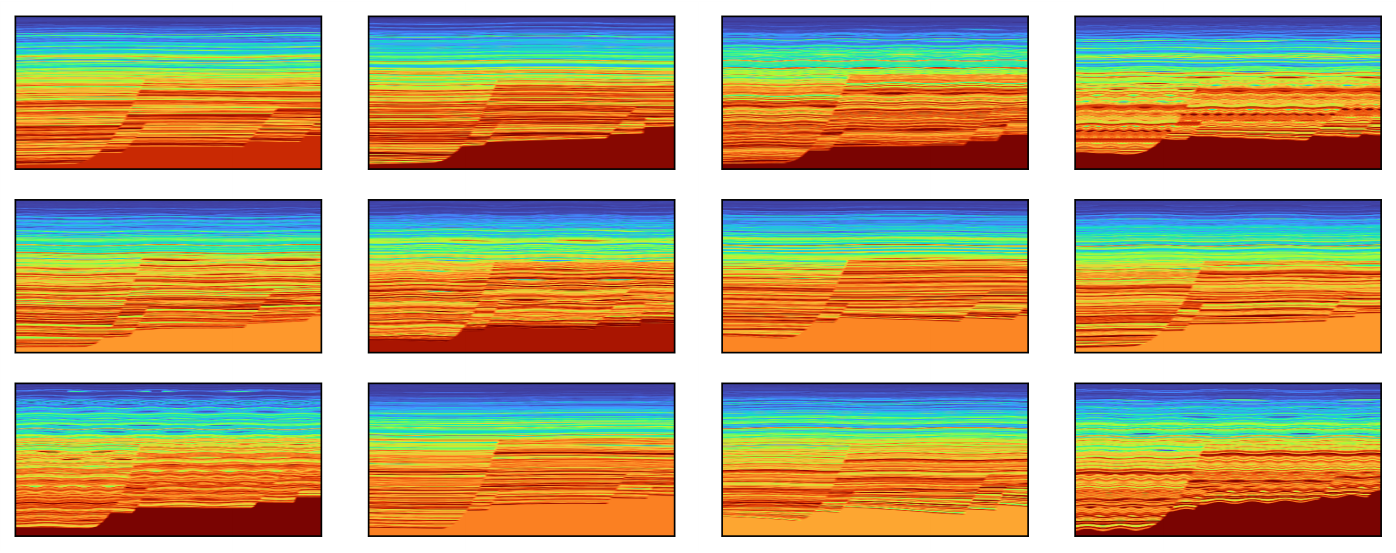}\label{fig:input1}
    \caption{The Otway model example: nine examples of the prepared velocity models for training.}
    \label{fig:trainvel_otway}
\end{figure}

\begin{figure*}[htpb!]
    \centering
    \subfloat[]{\includegraphics[width=0.75\columnwidth]{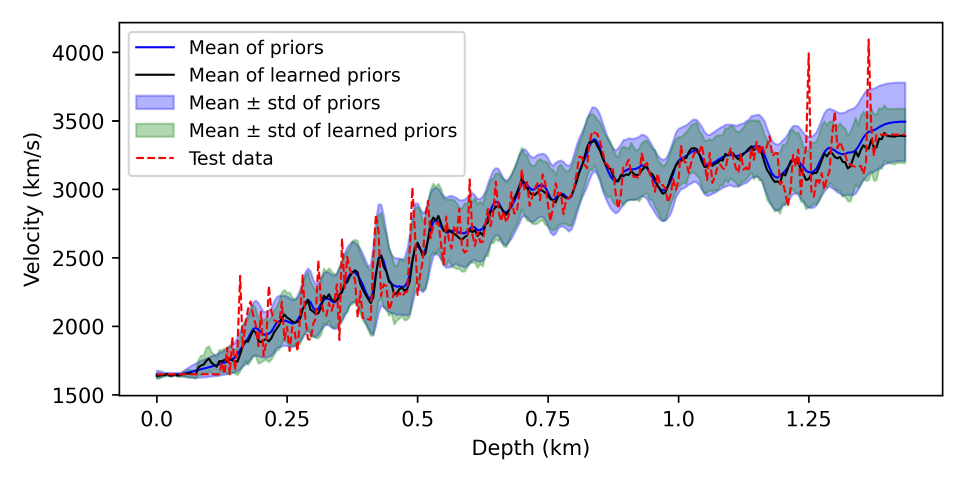}\label{fig:vel_dis055}}\\
    \subfloat[]{\includegraphics[width=0.75\columnwidth]{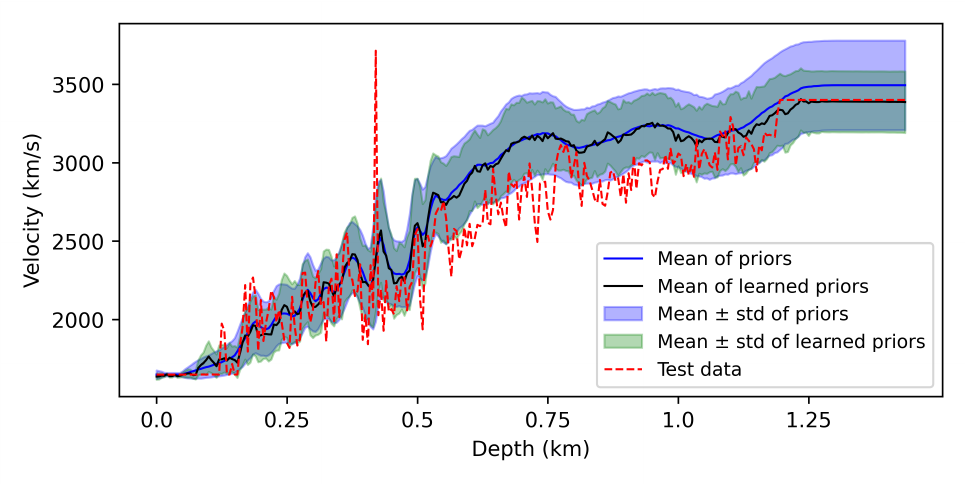}\label{fig:veldis18}}\\
    \caption{Comparison of original and learned prior distributions for velocity profiles at distance (a) 0.55 km and (b) 1.80 km. The original prior distribution, computed from 4,500 velocity model samples, is shown by the blue line (mean) and blue shaded region ($\pm\sigma$). The learned prior distribution, derived from velocity models generated by the trained diffusion model, is represented by the black line (mean) and green shaded region ($\pm\sigma$).}
    \label{fig:velpro}
\end{figure*}

\begin{figure*}[htpb!]
    \centering
    \includegraphics[width=0.60\columnwidth]{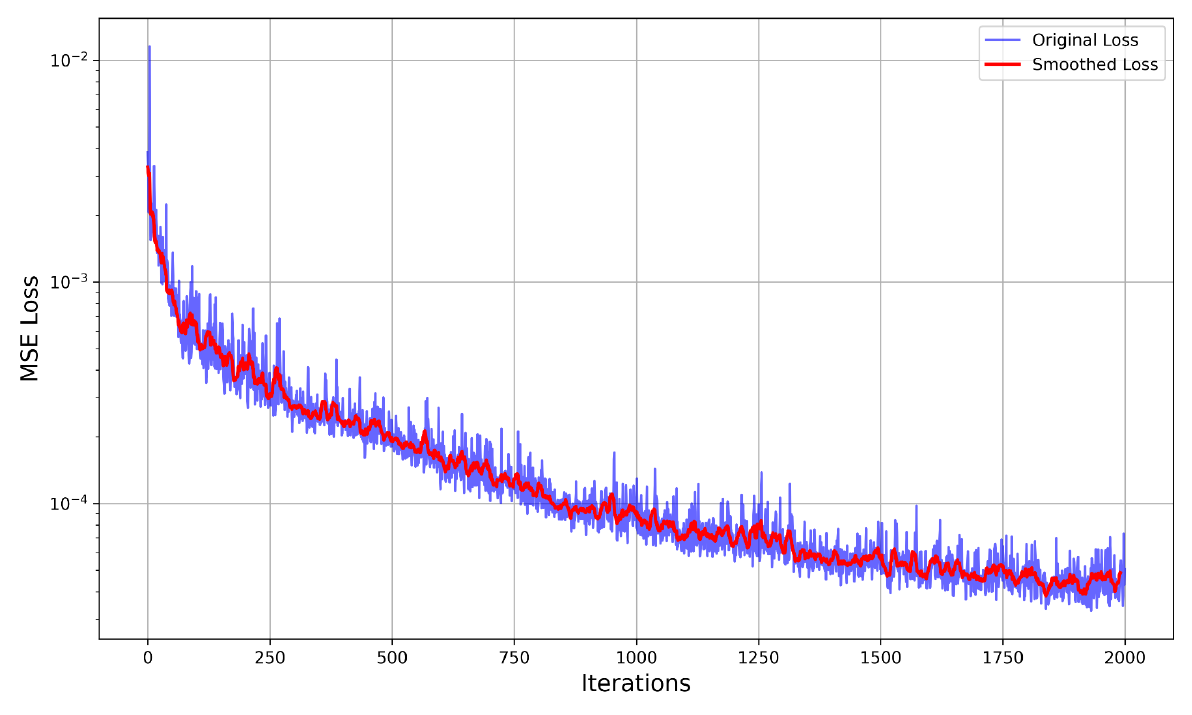}\label{fig:true}
    \caption{The convergence curve of the loss function in the fine-tuning process.}
    \label{fig:Invvel}
\end{figure*}

\begin{figure*}[htpb!]
    \centering
    \subfloat[]{\includegraphics[width=0.5\columnwidth]{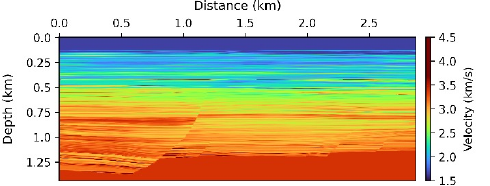}\label{fig:true}}
    \subfloat[]{\includegraphics[width=0.5\columnwidth]{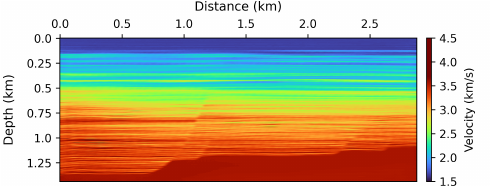}\label{fig:diffusioninv}}\\
    \subfloat[]{\includegraphics[width=0.5\columnwidth]{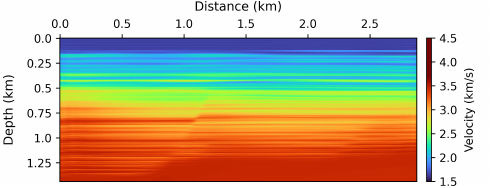}\label{fig:meanotway}}
    \subfloat[]{\includegraphics[width=0.5\columnwidth]
    {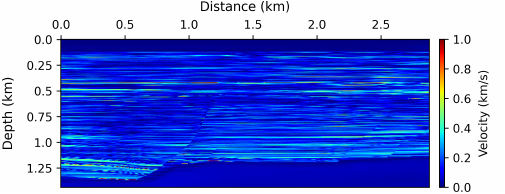}\label{fig:meanerr_otway}}\\
    \subfloat[]{\includegraphics[width=0.5\columnwidth]
    {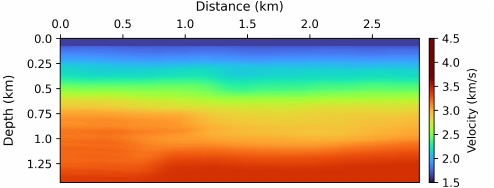}\label{fig:fwi_ini}}
    \subfloat[]{\includegraphics[width=0.5\columnwidth]{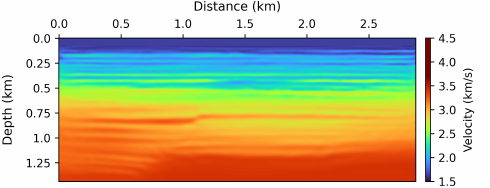}\label{fig:fwi_otw}} \\
    \caption{The Otway model example: (a) true velocity, (b) one example velocity sampling from the approximate posterior distribution, (c) the mean velocity of the approximate posterior distribution, (d) the velocity error between the mean velocity and the true one, (e) initial velocity for conventional FWI, and (f) the conventional FWI result. }
    \label{fig:Invvel}
\end{figure*}

\begin{figure*}[htpb!]
    \centering
    \subfloat[]{\includegraphics[width=0.5\columnwidth]{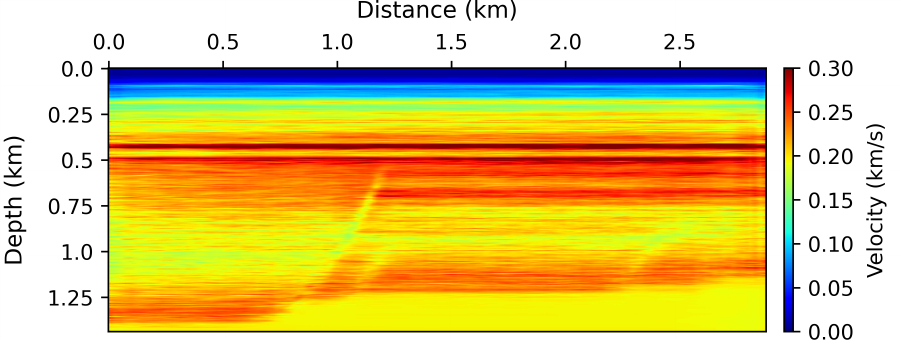}\label{fig:fwi}} 
    \subfloat[]{\includegraphics[width=0.5\columnwidth]
    {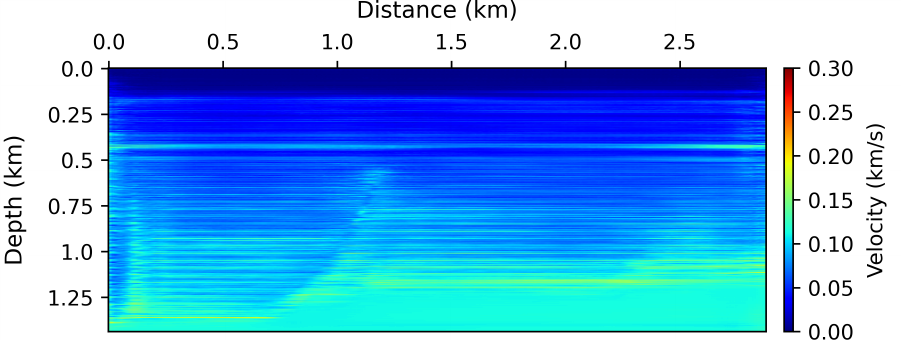}\label{fig:std}}\\

    \caption{The Otway model example: (a) the standard deviation of the learned prior distribution after pretraining, and (b) the standard deviation of the approximate posterior distribution after finetuning. }
    \label{fig:Invvel}
\end{figure*}

\begin{figure*}[htpb!]
    \centering
    \includegraphics[width=0.95\columnwidth]{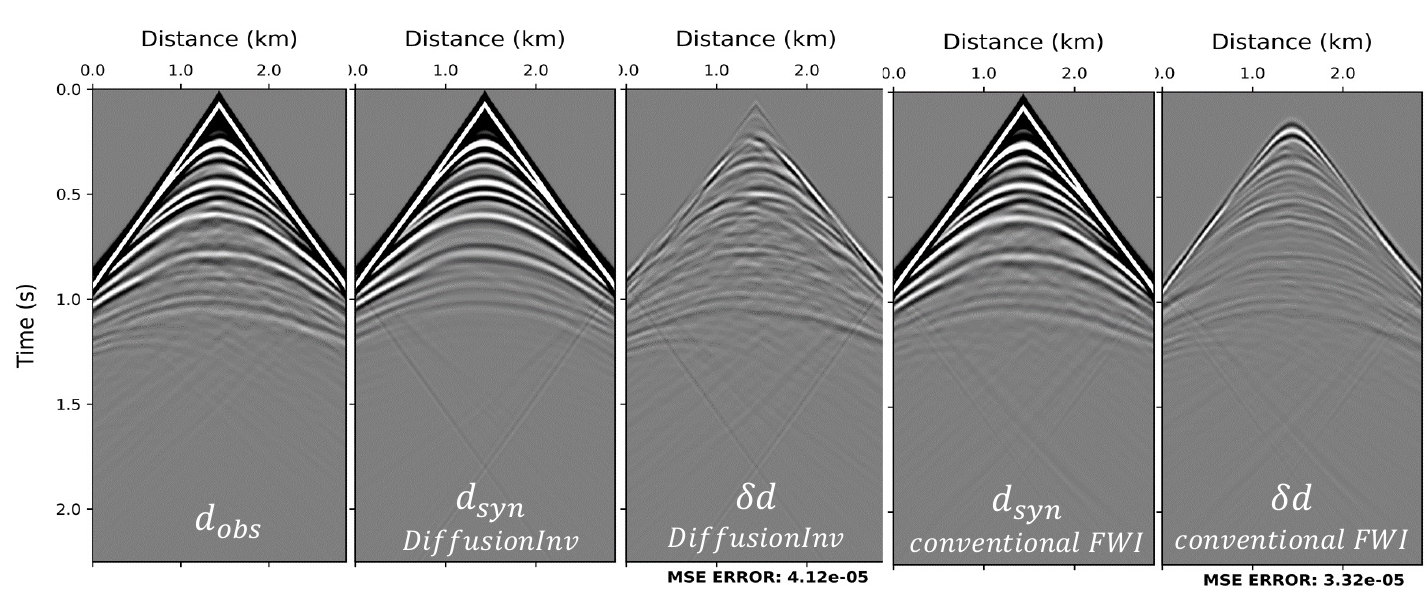}\label{fig:initial}\\

    \caption{The comparison between the observed and simulated shot gathers for the DiffusionInv mean velocity and conventional FWI result.}
    \label{fig:Invvel}
\end{figure*}


\begin{figure*}[htpb!]
    \centering
    \subfloat[]{\includegraphics[width=0.75\columnwidth]{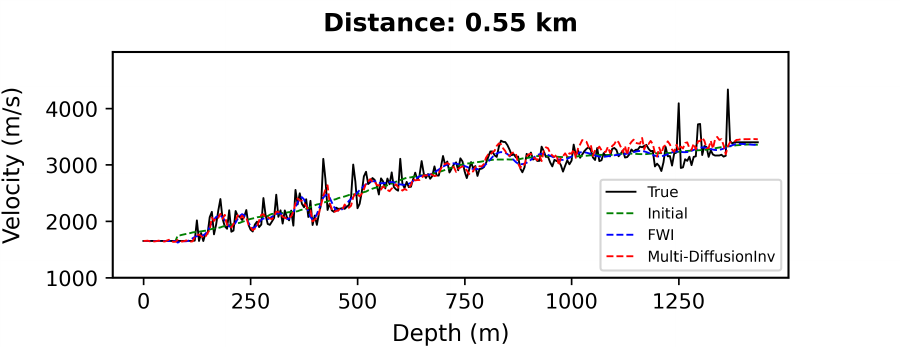}\label{fig:velpro_0d55}}\\
    \subfloat[]{\includegraphics[width=0.75\columnwidth]{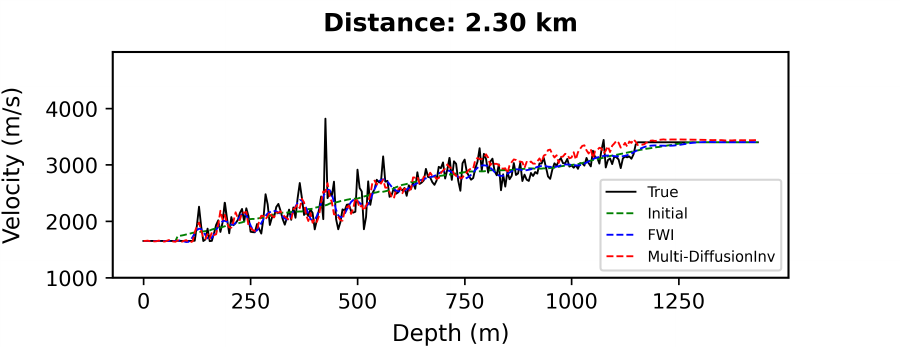}\label{fig:velpro_2d3}}\\

    \caption{The velocity profiles at distance (a) 0.55 km and (b) 2.30 km.}
    \label{fig:velpro}
\end{figure*}

\section{Discussion}
Both Bayesian Full Waveform Inversion (FWI) and diffusion models are built upon a probabilistic inference framework, and our proposed DiffusionInv method leverages the strengths of both approaches by integrating them into a unified inversion workflow. This DiffusionInv approach not only predicts subsurface velocity with high resolution but also obtains a reasonable estimate of the posterior distribution of the subsurface velocity model given the observed seismic data. The standard deviation of this posterior distribution is a crucial indicator for quantifying the uncertainty and reliability of the inversion result.   To recover the subsurface velocity that both satisfies prior model knowledge and fits the seismic observations, the DiffusionInv approach involves two main stages: prior-information pre-training and physics-informed fine-tuning. In the pretraining stage, a diffusion model is trained on a set of velocity models containing our prior knowledge of the subsurface plausible features. This process learns a prior distribution that enables the generation of velocity models with geological features similar to those observed in the training velocity models. Once the pretraining is complete, the diffusion model is fine-tuned to approximate the posterior distribution of the subsurface velocity by optimizing the data fitting between simulated and observed data. 

Traditional regularization methods are typically based on simple prior model assumptions, such as smoothness or sparsity. However, as exploration advances, we know more about the subsurface, such as well data, geological knowledge, and thus the prior model distributions can reflect such knowledge within a range of confidence. Diffusion models, as powerful generative models, are capable of learning and representing these complex prior distributions, making them well-suited for integrating geological knowledge into the inversion process. While data-driven FWI methods rely on extensive training datasets, their generalization across different geological scenarios remains a challenge. DiffusionInv mitigates this issue by incorporating physical constraints via the seismic wave equation, enabling it to adapt to various seismic acquisition geometries without requiring an extensive labeled dataset.

Prior knowledge of the subsurface model (e.g., geological structures, rock properties, or velocity-depth trends), typically derived from geological surveys, well logs, or existing seismic images and interpretations, plays a crucial role in guiding the construction of training velocity models. However, when reliable prior knowledge is limited, the velocity parameters should span a broader range of plausible values. Consequently, the diffusion model pretrained on these velocity models learns a prior distribution with a higher standard deviation, reflecting a broader search space, thus enhancing the model's capacity to explore geologically plausible velocity. However, it also increases computational demand during inversion, requiring more iterations to converge to geologically plausible solutions. Conversely, sufficient and reliable prior knowledge narrows the solution space, allowing the DiffusionInv approach to generate velocity models that satisfy both prior model information and observed seismic data with less iterations. This trade-off highlights the need to balance prior uncertainty (to prevent bias) with data-driven constraints (to ensure reliable convergence). 


A critical aspect of the DiffusionInv approach is achieving a balance between the influence of the learned prior and the physics-driven seismic data constraint. On one hand, the prior model distribution derived from the diffusion model provides geological constraints that steer the inversion toward geologically plausible solutions. On the other hand, the data fitting component ensures that the inverted model honors the physics captured in the seismic data. Over-reliance on the prior distribution may result in biased inversion results that overly reflect the characteristics of the training data, potentially overlooking the features present in the observed seismic data. Conversely, if the inversion relies too heavily on seismic data fitting, it may amplify noise and artifacts, especially in cases where the data is limited or contaminated by noise.

While the DiffusionInv approach demonstrates significant advantages in integrating prior model information and physical constraint of seismic data into FWI, along with providing a reliable estimate of the posterior distribution of subsurface velocity, it still faces the following challenges. The accuracy of the learned prior distribution depends on the representativeness of the training velocity models. If the training dataset lacks diversity or does not capture critical geological features, the inversion may struggle to generate reasonable velocity models. Besides, although the generative diffusion model enables efficient sampling, achieving optimal convergence during fine-tuning still requires up to 2000 iterations to reconcile physical constraint of seismic data with prior model information. Notably, traditional Bayesian inversion methods—such as Markov Chain Monte Carlo (MCMC)—often demand orders of magnitude more iterations due to their reliance on random sampling in high-dimensional parameter spaces. In our future work, we will further enhance the effectiveness of the proposed method by improving the learned prior information and the fine-tuning architecture.

\section{Conclusion}
We proposed the DiffusionInv method, which combines the generative diffusion model with FWI into a unified probabilistic inversion framework. This method can estimate the geologically plausible subsurface model that fits seismic observations by incorporating both the prior model information and physical constraint of seismic data into the inversion. More important, the proposed DiffusionInv approach can obtain the posterior probability distribution of subsurface velocity model given observed seismic data to quantitatively analyze the uncertainty of inversion. The numerical test results demonstrate that the DiffusionInv method recovered the velocity model with very high resolution beyond the resolution offered by conventional FWI.

\bibliographystyle{unsrt}  
\bibliography{DiffusionInv}

\end{document}